\documentclass[10pt,aps,pra,reprint,superscriptaddress]{revtex4-1}
\usepackage{graphicx}
\usepackage[colorlinks=true,linkcolor=blue,citecolor=blue,urlcolor=blue]{hyperref}
\usepackage[exponent-product=\cdot]{siunitx}

\makeatother
\begin{document}

\title{Topographic Effects on Ambient Dose Equivalent Rates from Radiocesium Fallout}

\author{Alex Malins}\email[Corresponding author: ]{malins.alex@jaea.go.jp}
\affiliation{Center for Computational Science \& e-Systems, Japan Atomic Energy Agency, 178-4-4 Wakashiba, Kashiwa-shi,
Chiba-ken, 277-0871, Japan}

\author{Masahiko Okumura}
\affiliation{Center for Computational Science \& e-Systems, Japan Atomic Energy Agency, 178-4-4 Wakashiba, Kashiwa-shi,
Chiba-ken, 277-0871, Japan}

\author{Masahiko Machida}
\affiliation{Center for Computational Science \& e-Systems, Japan Atomic Energy Agency, 178-4-4 Wakashiba, Kashiwa-shi,
Chiba-ken, 277-0871, Japan}

\author{Kimiaki Saito}
\affiliation{Fukushima Environmental Safety Center, Sector of Fukushima Research and Development, Japan Atomic Energy Agency, 2-2-2 Uchisaiwai-cho, Chiyoda,
Tokyo, 100-8577, Japan}

\date{\today}

\begin{abstract}
Land topography can affect air radiation dose rates by locating radiation sources closer to, or further from, detector locations when compared to perfectly flat terrain. Hills and slopes can also shield against the propagation of gamma rays. To understand the possible magnitude of topographic effects on air dose rates, this study presents calculations for ambient dose equivalent rates at a range of heights above the ground for varying land topographies. The geometries considered were angled ground at the intersection of two planar surfaces, which is a model for slopes neighboring flat land, and a simple conical geometry, representing settings from hilltops to valley bottoms. In each case the radiation source was radioactive cesium fallout, and the slope angle was varied systematically to determine the effect of topography on the air dose rate. Under the assumption of homogeneous fallout across the land surface, and for these geometries and detector locations, the dose rates at high altitudes are more strongly affected by the underlying land topography than those close to ground level. At a height of \SI{300}{\metre}, uneven topographies can lead to a \SI{50}{\percent} change in air dose rates compared to if the ground were uniformly flat. However, in practice the effect will more often than not be smaller than this, and heterogeneity in the source distribution is likely to be a more significant factor in determining local air dose rates.
\end{abstract}

\maketitle

\section{Introduction}

Surveys of gamma air radiation levels play an important role in quantifying radiological hazards if fallout from a nuclear accident or test contaminates an environment with radioactive material. The results typically quoted from radiation surveys are values of the ambient dose equivalent rate, $\dot{H}^*(10)$~[\si{\micro\sievert\per\hour}], at \SI{1}{\metre} above ground level, commonly referred to as the ``air dose rate''~\cite{Andoh2014}.

Data obtained from air dose rate surveys can be used to determine the type and quantity of radioactivity present in the environment. The radionuclides in the fallout can be identified by analyzing the energy spectrum of the gamma radiation at the detector~\cite{Torii2013,Tsuda2014}. The activity of radionuclides within the ground can then be estimated by employing conversion coefficients between \SI{1}{\metre} ambient dose equivalent rates and soil activity levels~\cite{Saito2014}.

With this knowledge, it is possible to calculate how air dose rates will change in future by performing decay calculations and modeling radionuclide migration~\cite{Kinase2014,Kitamura2014}. It is also possible to estimate external gamma radiation doses for exposed populations, and internal doses due to inhaling re-suspended activity or ingesting contaminated food-stuffs.

Fallout from the Fukushima Daiichi Nuclear Power Plant (FDNPP) accident in 2011 deposited radioactive material over wide areas of North East Japan. Data obtained from air dose rate surveys undertaken after the accident were primary inputs into the decision to evacuate highly contaminated regions in the early weeks after the accident~\cite{Lyons2012}. The data were also used in assessments of radiation doses received by the exposed populations~\cite{WHO2013,UNSCEAR2014}.

The air dose rate surveys employed various practices to gather measurements. In situ monitoring and vehicle surveys were conducted at ground level~\cite{Tanigaki2013,Saito2014a,Andoh2014}. Helicopters and fixed-wing aircraft were used to gather data covering wide areas from altitude~\cite{Lyons2012,Sanada2014}. The complementary programs have now provided coverage for much of Japan.

One limitation for converting between air dose rates and soil radioactivity levels is that the conversion factors are typically derived assuming perfectly flat land surfaces~\cite{Saito2014}. Uneven land topographies can alter air dose rates. For example, an ascending slope near to a detection point will result in a higher air dose rate compared to if the ground were flat, as, assuming the total source activity and horizontal distribution are unchanged, the radioactive source is closer to the detector in the former case. The converse is true for slopes descending away from detectors; the air dose rate will be lower than if there were flat ground as the source is further from the detector~\cite{Schwarz1992}.

Topographic effects can also cause uncertainty in converting measurements taken at altitude in aircraft radiation surveys to ground-level values. For instance, the ground-level air dose rate can be estimated from a high altitude measurement by using a conversion factor between the \SI{1}{\metre} ambient dose equivalent rate measured at a calibration site to a detector response measured in an aircraft at various elevations above that site~\cite{Sanada2014,Hendricks1999}. As the calibration sites are typically located in flat areas to allow vehicle access to make the ground-level measurement, it is not clear \textit{a priori} that the derived conversion factors will also be applicable if the survey goes on to cover mountainous areas~\cite{Lyons2012,Sanada2014,Sanada2015}.

Previously Allyson investigated the effect of topography on unscattered photon fluence rates at various altitudes using an analytical method~\cite{Allyson1994}. He considered \textsuperscript{137}Cs fallout on the soil surface in valleys and on mountain ridges. The largest effects were seen for steep topographic features and for high detector heights. Saito carried out Monte Carlo calculations for various different land topographies, but with natural radionuclides, and found deviations in air kerma rates of up to \SI{50}{\percent}~\cite{Saito2000}. Schwarz~{et~al.}~introduced a method for correcting airborne gamma measurements for the perturbations induced by large scale topographic features~\cite{Schwarz1992}.

Satoh \textit{et al.}~performed Monte Carlo calculations for ambient dose equivalent rates from radiocesium fallout for smaller topographic features, in particular \SI{5}{\metre} slopes up to \SI{40}{\degree} in inclination~\cite{Satoh2014}. Slopes of this size are similar to engineered slopes and embankments found within Fukushima Prefecture, and they found that topography could lead to a \SI{20}{\percent} enhancement in ground-level dose rates.

The mountainous region lying to the west of the Fukushima Daiichi site, known as the Nakadori area of Fukushima Prefecture, suffered large deposits of radioactive \textsuperscript{134}Cs and \textsuperscript{137}Cs. This situation contrasts with the 1986 Chernobyl accident, where the most serious deposits of fallout were over relatively flat areas of Ukraine and Belarus in comparison. Therefore we ask how large the effect of wide area topographical features, such as mountains and valleys, can be on air dose rates. We consider the uncertainty induced by topographic effects when converting \SI{1}{\metre} air dose rates into soil activity levels. We also consider air dose rates at higher elevations above the ground, to analyze the topographic uncertainty when converting radiation measurements taken within aircraft to ground-level values.

\section{\label{sec:Methods}Methods}

We considered the effect of topography on air dose rates at heights $h$~[\si{\metre}] above land surface, from \SI{1}{\centi\metre} up to \SI{300}{\metre}. The higher altitudes cover ranges typically sampled in aircraft surveys. Note \SI{1}{\metre} above ground level is the standard height for quoting ambient dose equivalent rates in the event of a nuclear accident~\cite{TECDOC1162}.

\begin{figure}
	\centering
	\includegraphics[width=0.45\textwidth]{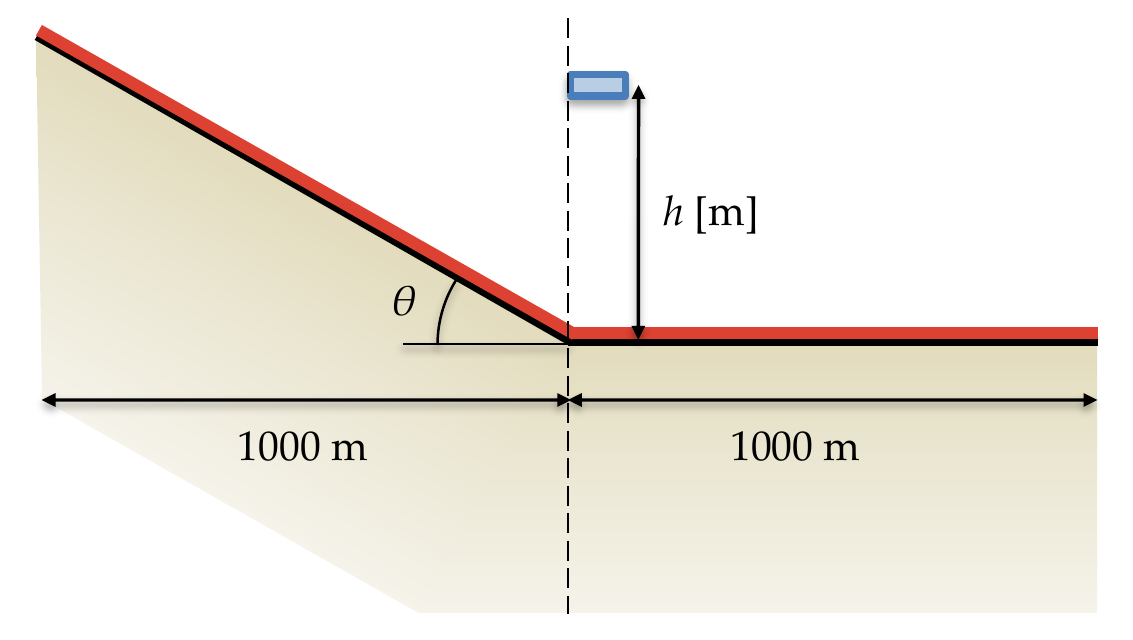}
	\caption{\label{fig:angled_ground}Angled ground geometry. The detector is the blue volume. Fallout is indicated as a red layer on the land surface, and soil is beige.}
\end{figure}

Two idealized geometries were simulated using the Monte Carlo radiation transport code PHITS~\cite{Sato2013}. The first was angled ground at the intersection of two planar surfaces (fig.~\ref{fig:angled_ground}). The geometry consisted of soil and air volumes within a \SI{1}{\kilo\metre} radius bounding cylinder. Flat ground changes abruptly to a uniformly angled slope along one direction. The height of the air in the simulation above the flat land was \SI{400}{\metre}, and the minimum depth of soil at all locations was \SI{1}{\metre}.

Gamma tracks were detected in \SI{10}{\metre} length and breadth cuboids above the flat ground adjacent to the slope. To generate sufficient statistics, the vertical thickness of the detector volumes varied between \SI{0.5}{\centi\metre} at ground level, to \SI{2}{\metre} at the highest altitudes.

\begin{figure}
	\centering
	\includegraphics[width=0.45\textwidth]{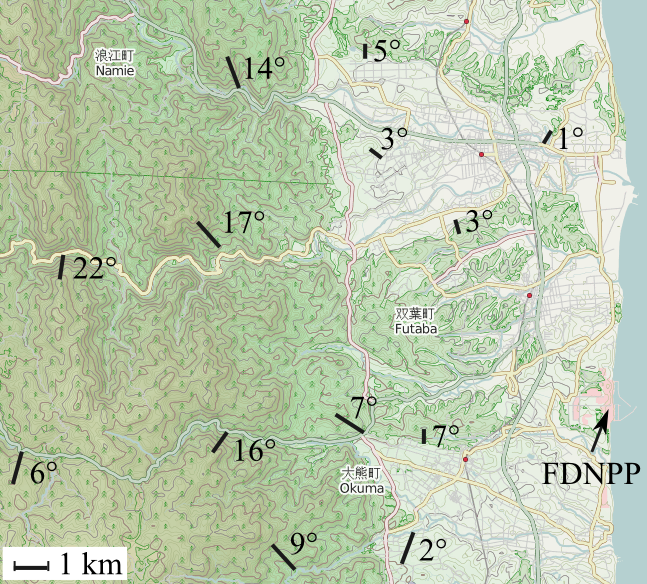}
	\caption{\label{fig:map_marks}Land topography near to FDNPP. Marked are typical land inclination angles for the area, as measured over \SI{500}{\metre} to \SI{1}{\kilo\metre} distances shown by black bars. Map tile Copyright Thunderforest, map data Copyright OpenStreetMap contributors (2014).}
	
\end{figure}

\begin{figure*}
	\centering
	\includegraphics[width=0.95\textwidth]{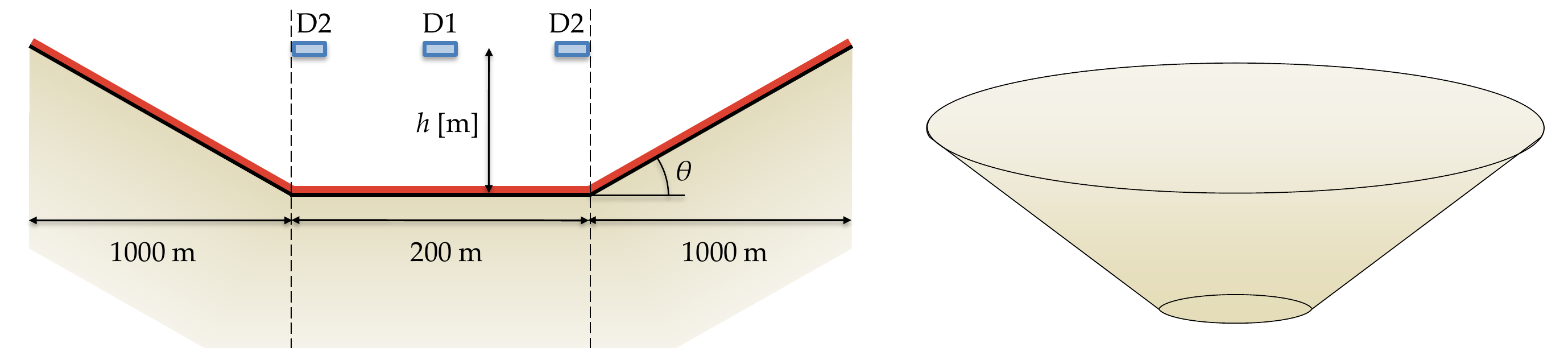}
	\caption{\label{fig:conical}Conical geometry with two detector volumes -- D1 and D2.}
\end{figure*}

\begin{figure*}
	\centering
	\includegraphics[width=0.95\textwidth]{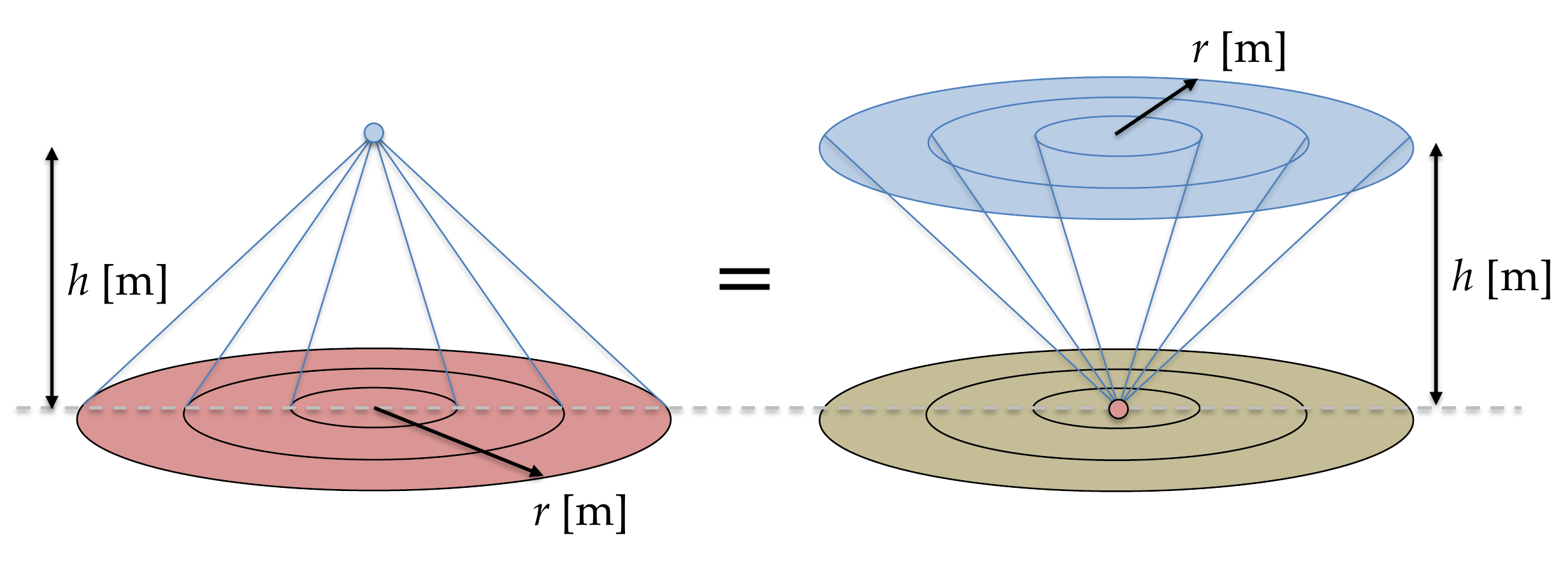}
	\caption{\label{fig:flat_saito_method}Simulation method for the flat ground model. Red indicates radiocesium (the source region), blue detector volumes, and beige uncontaminated soil.}
\end{figure*}

The angle of inclination $\theta$ [\si{\degree}] of the slope was systematically varied between \SI{-20}{\degree} and \SI{20}{\degree}. This range of land gradients is characteristic of the urban and mountainous areas contaminated with radioactive cesium in Fukushima Prefecture, as can be seen from the map in fig.~\ref{fig:map_marks} that covers the area immediately west of FDNPP. Positive angles indicate slopes that ascend up from the flat ground, and negative angles the converse. In the simulations the angle $\theta=0$ corresponds to uniformly flat terrain.

The second geometry considered was a conical geometry. In this geometry, slopes ascend (or descend) away from a circular area of flat ground, radius \SI{100}{\metre}, in the center of the simulation (fig.~\ref{fig:conical}). This geometry approximates detector locations above hilltops and valley bottoms.

Photon detection took place in cylinders above the center of the flat area (denoted D1), and annular cylinders above the flat ground bordering the slope (D2). The radius of the D1 cylinders was \SI{10}{\metre}, and the horizontal thickness of the D2 annular cylinders was also \SI{10}{\metre}. The vertical thickness of the detector volumes varied between \SI{0.5}{\centi\metre} and \SI{2}{\metre}, depending on the altitude of the dose rate point. The slope angle was varied between \SI{-20}{\degree} and \SI{20}{\degree}, and other parameters for this geometry were similar to the angled ground simulations.

The results for these topographies were compared against a reference case of uniformly flat ground. In this case there is $x$-$y$ invariance in the geometry and it is possible to rescale the source and detector volumes to dramatically improve the computational efficiency of the simulation~\footnote{As long as the simulation space is sufficiently large that edge effects induced by the absorbing boundaries are negligible. We show this is the case for our simulations by checking convergence of the air dose rate to the infinite half-space limit.}  (see fig.~\ref{fig:flat_saito_method} and ref.~\cite{Namito2012}). This method also offers the benefit that it allows the contribution of different areas of the source on the ground to the overall air dose rate to be calculated. In particular we calculated the contribution from annular rings of source, between inner and outer radii of $r$~[\si{\metre}] and $r+1$ respectively, in order to check the finite size scaling of the results for the air dose rates.

All models consisted of two materials -- air and soil. The atomic composition of these substances matched the values given in ref.~\cite{Eckerman1993}. The density of air was \SI{1.2e-3}{\gram\per\cubic\centi\metre} and soil was \SI{1.6}{\gram\per\cubic\centi\metre}.

The radiation source was photons with either the \textsuperscript{134}Cs or \textsuperscript{137}Cs decay energy spectrum. The spectra were drawn from NuDat2, which is based on  the Evaluated Nuclear Structure Data File (ENSDF)~\cite{NuDat2}.

The radiation source was a thin layer of contamination on the surface of the ground, or at \SI{1}{\centi\metre} depth within the soil. The former case equates to a relaxation mass per unit area of $\beta=0$~[\si{\gram\per\square\centi\metre}].

Note the radiocesium activity per unit area of land when projected onto a horizontal plane was constant for all simulations. This allows fair comparison of the effect of topography on dose rates between the different slope inclinations. In other words the total radioactivity of cesium source was constant for all slope inclinations in the angled ground and conical simulations.

\begin{figure*}
	\centering
	\includegraphics[width=0.95\textwidth]{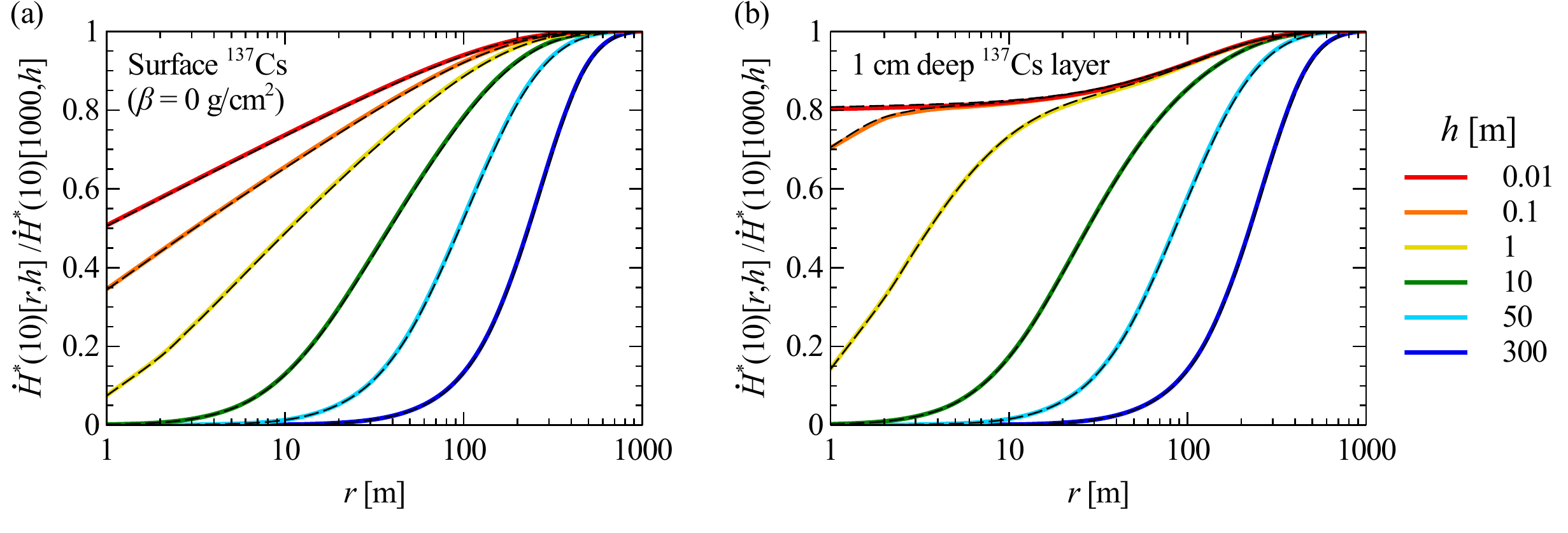}
	\caption{\label{fig:finite_size}The fractional yield of the infinite half-space ambient dose equivalent rate for a radius $r$ source on uniformly flat ground. Colored lines are for \textsuperscript{137}Cs, and superimposed dashed lines for \textsuperscript{134}Cs.}
\end{figure*}

Ambient dose equivalent rates were calculated by multiplying the flux detected at each height by conversion factors for $H^*(10)$ from ICRP~74~\cite{ICRP74}. The flux was tallied within \SI{0.5}{\kilo\electronvolt} wide energy bins and logarithmic interpolation was used between the point $H^*(10)$ conversion factors.

Note that none of the models studied here include any fine detail for the land structure or use, such as buildings, tarmac, surface water, vegetation or local variations in topography.

\section{\label{sec:Results}Results}

\subsection{\label{ssec:Field_of_View}Field of View}

The fields of view for airborne gamma measurements can extend to hundreds of meters in radius on the ground~\cite{Allyson1994,Tyler1996}, so first we demonstrate that our simulation spaces were sufficiently large to avoid to finite size effects in the calculated dose rates. Fig.~\ref{fig:finite_size}(a) displays the fractional air dose rate for limited radii sources compared to the asymptotic limit for an infinite half-space. These simulations were of a thin radiocesium layer on the surface of uniformly flat land.

The higher the detector elevation, the wider the field of view of the detector. At \SI{300}{\metre} elevation the dose rate from a \SI{900}{\metre} radius source is \SI{99.8}{\percent} of the value for a \SI{1}{\kilo\metre} radius source. This indicates that there is no significant contribution to the total dose rate from any source located beyond \SI{1}{\kilo\metre}.

The field of view narrows when the radiocesium layer is within the soil pack (fig.~\ref{fig:finite_size}(b)). The effect is most pronounced for detectors close to the land surface. This is because lower elevation detectors have wider fields of view with respect to detector height, i.e.~larger solid angles, hence the photon path length within soil is longer~\cite{Tyler1994}.

For both surface and \SI{1}{\centi\metre} deep soil layers, the fields of view coincide for \textsuperscript{134}Cs and \textsuperscript{137}Cs contamination. This is because the \textsuperscript{134/137}Cs decay photons that make the dominant contribution to the dose rate are emitted within the energy range \num{0.6} to \SI{0.8}{\mega\electronvolt}, and so have comparable mean free paths in air and soil. For the angled ground and conical geometries, we simulated only \textsuperscript{137}Cs sources and infer that the results for \textsuperscript{134}Cs will be quantitatively similar.

\subsection{\label{ssec:Angled_Ground_Geometry}Angled Ground Geometry}

\begin{figure*}
	\centering
	\includegraphics[width=0.95\textwidth]{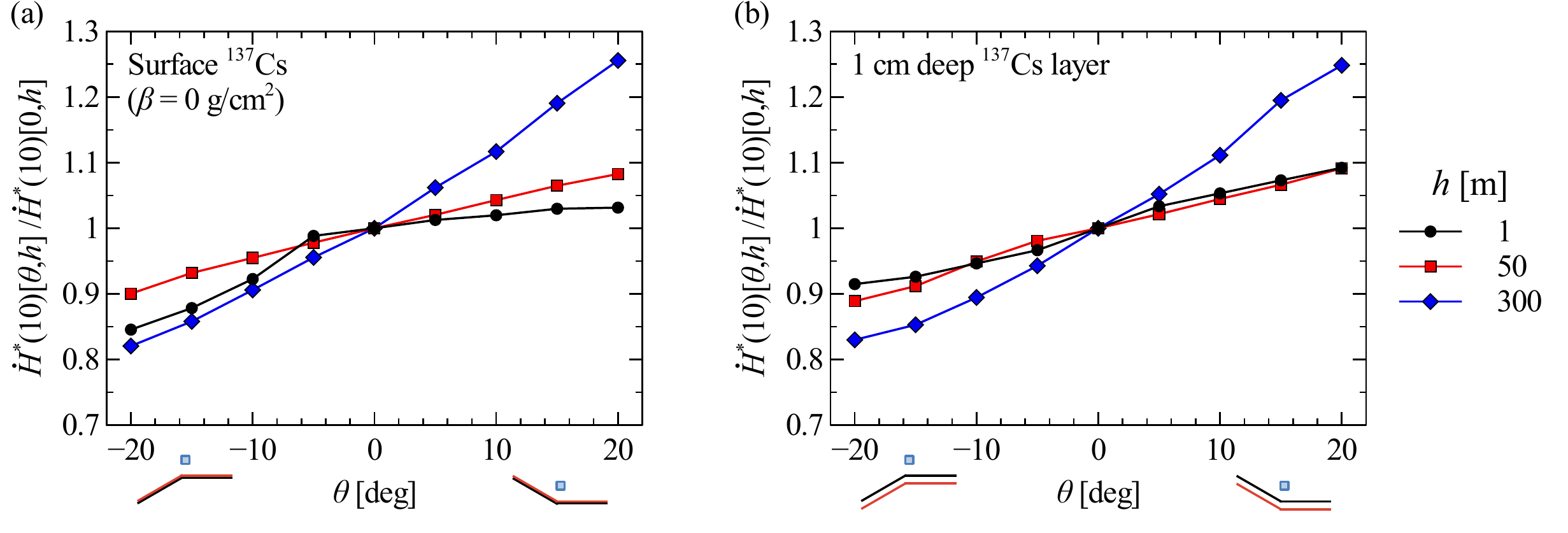}
	\caption{\label{fig:angle}Relative change in ambient dose equivalent rates for different land inclination angles in the angled ground geometry. Schematics of the geometries, sources and detector locations for positive/negative $\theta$ are shown below the abscissa.}
\end{figure*}

Slopes ascending up from flat ground result in increased dose rates, as the source is closer to the detector than for uniformly flat ground (fig.~\ref{fig:angle}, positive $\theta$). Conversely, slopes descending away from the detector result in lower dose rates, because the radiation has to travel further to reach the detector.

The effect for \SI{1}{\metre} high detectors is relatively small when the \textsuperscript{137}Cs fallout is on the soil surface (fig.~\ref{fig:angle}(a)). The maximum increase in the dose rate for ascending slopes up to \SI{20}{\degree} is \SI{3}{\percent}. A change in behavior is visible for the descending slopes between \SI{-5}{\degree} and \SI{-10}{\degree} inclinations. The dose rates decrease faster relative to the flat ground case compared to the increase for positive slope inclinations. This occurs because, for sufficiently large negative inclinations, the soil shields the direct path between radiation originating on the slope and part of the detector volume.

For higher altitude detectors the effect of the topography on dose rates is generally stronger, particularly at \SI{300}{\metre} elevation where the dose rates change by up to \SI{26}{\percent} relative to the flat ground case. However, for the higher elevations, the geometries mean that slopes descending away from flat ground do not shield the source from the detector. For \num{50} and \SI{300}{\metre} detector elevations there is little difference between radiocesium layers on the soil surface and at \SI{1}{\centi\metre} deep (c.f.~\ref{fig:angle}(a) and (b)).

The effect of topography for \SI{1}{\metre} detectors is slightly stronger for ascending slopes when the contamination layer is within the soil than when it is on the surface. Conversely for descending slopes, there is a smaller effect of topography as the narrower field of view means that attenuation by the slope itself is of lesser importance.

A limitation of these results is that they only apply for the simulated detector locations. It is possible to determine qualitatively how the dose rates would change at different positions. If the detector moves away from the slope and towards the flat region, the limiting dose rate is identical to the dose rate for uniformly flat ground.

\begin{figure}
	\centering
	\includegraphics[width=0.45\textwidth]{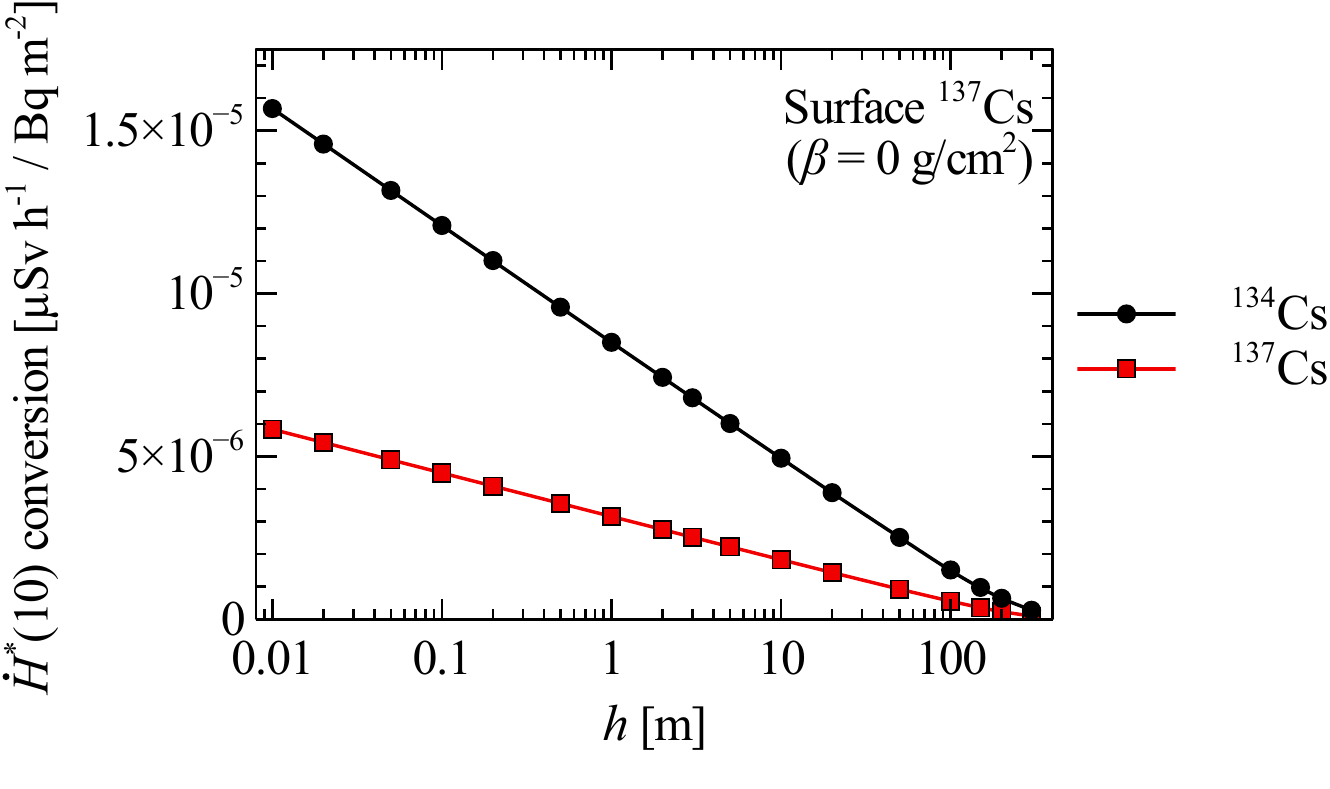}
	\caption{\label{fig:flat_height}Ambient dose equivalent rate conversion factors as a function of elevation above perfectly flat ground.}
\end{figure}

Moving in the opposite direction, i.e.~up or down the slope, the limiting value is the dose rate at height $h\cos{\theta}$ above a uniformly flat surface. Thus there is a change in the effective height of the detector. This is a purely geometrical effect attributable to the difference between the vertical height above the slope and the shortest distance to it. 

To elucidate the consequence of this effect, fig.~\ref{fig:flat_height} shows $\dot{H}^*(10)$ conversion factors per unit \textsuperscript{134}Cs and \textsuperscript{137}Cs surface activity for various elevations above flat ground. The ambient dose equivalent rate shows a logarithmic reduction with height. Therefore, the effective reduction of the detector altitude by $\cos{\theta}$ results only in a small increase in the air dose rate.

\subsection{\label{ssec:Conical_Geometry}Conical Geometry}

\begin{figure*}
	\centering
	\includegraphics[width=0.95\textwidth]{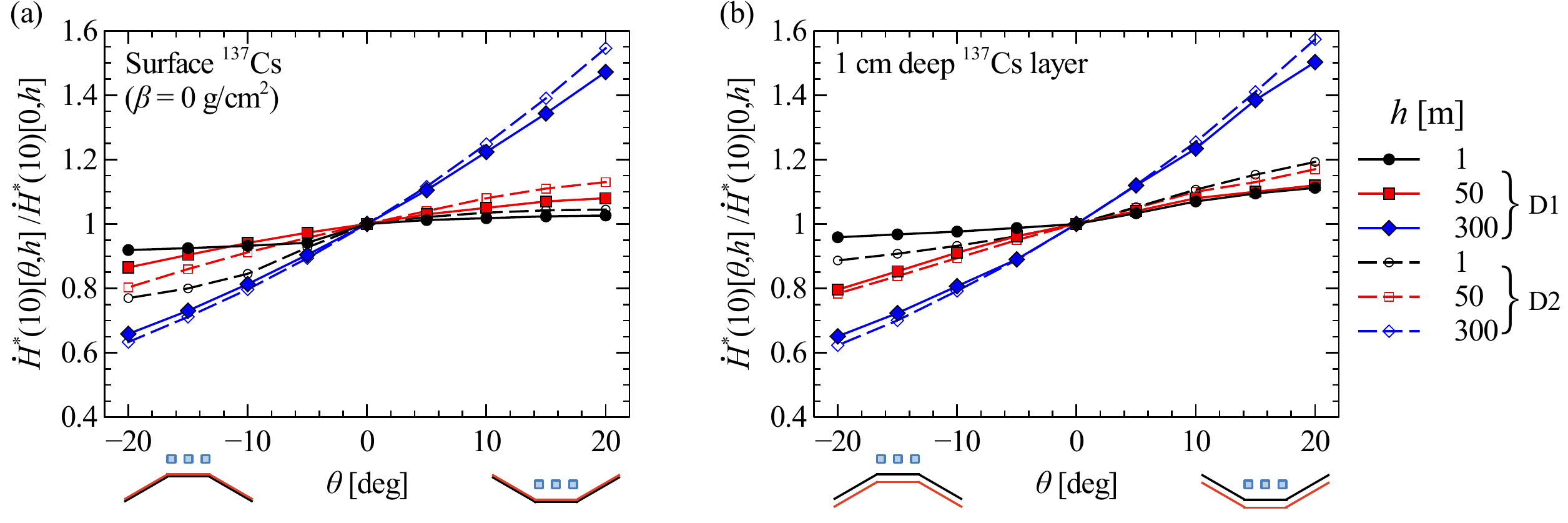}
	\caption{\label{fig:conical_graph}Relative change in ambient dose equivalent rate with slope inclination for the conical geometry.}
\end{figure*}

The second type of land geometries considered were conical slopes surrounding flat land. The inclination angle of the slopes was systematically varied, and the air doses rates compared against the uniformly flat ground reference case, fig.~\ref{fig:conical_graph}.

The trend of the data is similar to that for the angled ground geometry, however the magnitude of the effect is stronger as slopes now surround the detectors. Slopes ascending near to a detector result in higher air dose rates, and slopes descending away lead to smaller dose rates. The effect is most pronounced for the largest slope inclinations, and for detector locations D2 that are adjacent to the slope. The dose rates differ by more than \SI{35}{\percent} for detectors at \SI{300}{\metre} elevation and large inclination angles when compared to perfectly flat land.

There is comparatively little difference between the two different source locations for \num{50} and \SI{300}{\metre} high detectors, c.f.~figs.~\ref{fig:conical_graph}(a) and (b). For the \SI{1}{\metre} ambient dose equivalent rates, there are slight variations between the results for the surface and \SI{1}{\centi\metre} deep source layers. These variations are attributable to the different fields of view and consequentially the relative effect of bringing the source closer to or further from the detector by changing the slope inclination. However, the magnitude of the effect is always small, and the dose rate is always within \SI{23}{\percent} of the dose rate for perfectly flat land.

\section{\label{sec:Discussion}Discussion}

\subsection{\label{ssec:Ground-Level_Air_Dose_Rates}Ground-Level Air Dose Rates}

The simulations were based on large topographic features, such as mountains and valleys, and the studied land inclination angles varied between \SI{-20}{\degree} and \SI{20}{\degree}. There are only small deviations in the \SI{1}{\metre} ambient dose equivalent rate compared to the flat ground case for this inclination angle range. Larger deviations are possible for smaller but steeper topographic features such as engineered slopes~\cite{Satoh2014}.

The difference between the air dose rates for uneven and perfectly flat ground sets the uncertainty due to topography when converting \SI{1}{\metre} dose rates into soil activity levels using standard conversion factors~\cite{Saito2014}. If the air dose rate is higher than otherwise would be the case for flat land due to the presence of ascending slopes near to the detector, the estimate for radiocesium concentration within the soil will be an overestimate. The soil activity will correspondingly be underestimated if slopes descend away from a detector location.

These results apply under the assumption of spatially homogeneous radiocesium fallout. In practice there can be large spatially variability in soil activity levels over relatively small areas~\cite{Tyler1996}. As the field of view for \SI{1}{\metre} ambient dose equivalent rates is on the order of \num{10} to \SI{100}{\metre}, uncertainty in predictions for soil activity is more likely to be due to spatial heterogeneity of contamination than topographic effects.

\subsection{\label{ssec:Airborne_Dose_Rate_Surveys}Airborne Dose Rate Surveys}

The effect of topography on air radiation dose rates was generally stronger for higher altitude detector locations, particularly at \SI{300}{\metre}, than for detectors close to the land surface. If converting between aerial gamma measurements and ground-level dose rates using a constant factor derived for flat land~\cite{Hendricks1999}, the uncertainty for uneven topographies may be as large as \SI{60}{\percent}. However, as this represents the largest deviation seen in the simulation results, the topographic uncertainty is likely to be smaller than this value in practice. 

The effect of the topographic uncertainty on the ground-level predictions from airborne surveys employing conversion factors suited for flat land is as follows. Predictions for ground-level dose rates near hilltops will be underestimates, and correspondingly predictions made for near valley bottoms will be overestimates~\cite{Allyson1994}.

\section{\label{sec:Conclusions}Conclusions}

For the geometries and detector locations studied here, the air dose rates for varying land topographies are always within \SI{60}{\percent} of their values for perfectly flat land. At detector heights near to ground level, the effect of large topographic features on air dose rates was small. Other factors, such as spatial heterogeneity of the contamination, land use, buildings and decontamination efforts, will be more important sources of uncertainty in calculations for soil activity levels based on measured dose rates. 

The largest effects of topographic uncertainty were seen for the high altitude detectors. Wide-area topographic features, such as mountains or valleys, could therefore be significant sources of uncertainty when interpreting the results from airborne dose rate surveys. Mountainous areas are also the regions where airborne gamma surveys are most powerful, as it is more difficult to access mountainous terrain at ground level to perform radiation surveys.

It is worth noting that comparable levels of topographic uncertainty were found for \SI{50}{\metre} elevation and ground-level detectors. Unmanned aerial vehicles can operate at lower altitudes than traditional fixed-wing aircraft or helicopters~\cite{MacFarlane2014,Sanada2015}, and therefore drone surveys will be less susceptible to topographic uncertainties in their processed output.

\begin{acknowledgments}

We acknowledge fruitful discussions with Y. Hasegawa, S. Nakayama, K. Niita, Y. Sanada, D. Sanderson and M. Yoneya. Calculations were performed on JAEA's BX900 supercomputer.

\end{acknowledgments}

\bibliography{sna15_conference_paper_formatted}

\end{document}